\documentclass{article}

\usepackage[utf8]{inputenc}
\usepackage{authblk}

\usepackage{amssymb}
\usepackage{amsthm}
\usepackage{parskip}
\usepackage{float}
\usepackage{booktabs}
\usepackage{blindtext}
\usepackage[english]{babel}
\usepackage{csquotes}

\usepackage{amsbsy,amsmath}
\usepackage{textcomp}
\usepackage{animate}
\usepackage[
	backend=biber,
	citestyle=apa,
	bibstyle=apa,
	maxcitenames=2,
	maxbibnames=99]{biblatex}
\DeclareNameAlias{sortname}{family-given}
\usepackage[colorlinks=true, allcolors=blue]{hyperref}

\usepackage{graphicx}
\graphicspath{ {./} }
\usepackage{caption}
\usepackage{subcaption}

\usepackage{amsmath}
\usepackage[T1]{fontenc}
\usepackage{cancel}
\usepackage{mathrsfs}

\date{October 2022}

\title{Wavefield reconstruction inversion modelling of Marchenko focusing functions}
\author[1,2]{Ruhul F. Hajjaj\footnote{~Email: eerfi@leeds.ac.uk. This document constitutes a summary of results in the PhD Transfer Report of R.F. Hajjaj, successfully defended on October 19th, 2022.}}
\author[1]{Sjoerd A.L. de Ridder}
\author[1]{Philip W. Livermore}
\author[3]{Matteo Ravasi}
\affil[1]{University of Leeds, Leeds, United Kingdom.}
\affil[2]{Institut Teknologi Sumatera, Lampung, Indonesia.}
\affil[3]{King Abdullah University of Science and Technology, Thuwal, Saudi Arabia.}

\begin{document}

\maketitle

\begin{abstract}
    Marchenko focusing functions are in their essence wavefields that satisfy the wave equation subject to a set of boundary, initial, and focusing conditions. Here, we show how Marchenko focusing functions can be modeled by finding the solution to a wavefield reconstruction inversion problem. Our solution yields all elements of the focusing function including evanescent, refracted, and trapped waves (tunneling). Our examples indicate that focusing function solutions in higher dimensions are however challenging to compute by numerical means due to the appearance of strong evanescent waves.
\end{abstract}

\section{Introduction}
Focusing functions are a new concept in wavefield propagation that lies at the foundations of various Marchenko imaging schemes. They enable Green's function retrieval inside a medium from a single-sided (surface based) recording of seismic reflection data. The Marchenko equations provide a relationship between the Green's function in the interior of a medium ($G_D$) and the reflection data at the surface ($R$) via the focusing functions ($f$) (\cite{Burridge80}; \cite{Broggini12}; \cite{Wapenaar14}; \cite{Wapenaar21a}). Conventionally, these kinds of focusing functions are estimated from the reflection data in a data-driven manner by means of Neumann series of direct inversion, e.g. (\cite{Broggini14}; \cite{vanderneut15}; \cite{Vargas21}). Even though focusing functions are solutions of the wave equation themselves, they are traditionally thought of more as a mathematical than a physical entity (\cite{Filho18}). To our knowledge, there are no theoretical proofs available for model-based validations in dimension higher than 1D. 

Marchenko focusing function modelling can be done using full-wavefield propagation methods when a detailed subsurface model is available. \cite{Elison21} use the two-way wavefield extrapolation method to model focusing functions (\cite{Kosloff83}; \cite{Wapenaar86}). Focusing state is achieved by incorporating the focusing boundary condition that is defined for the pressure (a delta function, or its spatio-temporal bandlimited version) and the particle velocity field (i.e. the vertical derivative of the delta function) (\cite{Wapenaar93}). The method eliminates evanescent and downward propagating waves at each integration step e.g. using spectral projectors (\cite{Sandberg2009}). This approach had earlier been indicated by \cite{Becker16} and \cite{Wapenaar17}. Meanwhile, focusing functions will not be able to perfectly focus the wavefield as long as evanescent waves are not compensated (\cite{Wapenaar21}).

In this study, we follow a model-driven approach to compute focusing functions. We define focusing function modelling as a full-wavefield reconstruction inverse problem in actual (non-truncated) media. In this way, functions can be generalised to accommodate full-wavefields propagation that can ultimately lead to more general Marchenko schemes, with the ability to accurately image steep flanks and to account for evanescent and refracted waves.

\section{Algorithm}

\subsection{Defining the focusing functions}
A focusing wavefield is a solution to the wave equation that forms a focus in space on a horizontal plane at a given focal depth (i.e. the focal plane). We define a focusing wavefield as the solution to a partial differential equation (PDE) and a set of boundary, initial, final, and focusing conditions, in and on a domain. The spatial boundaries of the domain are open (i.e. we impose no boundary condition) and the wavefield has support throughout the domain, with the exception of the focal plane, where the support is confined to the focal point. The focusing wavefields satisfy the following conditions:
\begin{eqnarray}
    \label{eq:FC1}
    \left[\rho(\mathbf{x})\nabla^T\frac{1}{\rho(\mathbf{x})}\nabla - \frac{1}{c^2(\mathbf{x})}\frac{\partial^2}{\partial t^2}\right]F_\pm(\mathbf{x},\mathbf{x}_f,t)\hspace{-2mm}&=&\hspace{-2mm}0, \\
    \label{eq:FC2}
F_\pm(\mathbf{x},\mathbf{x}_f,t)\hspace{-2mm}&=&\hspace{-2mm}0,\ \mathrm{for}\ t\downarrow-\infty,\\
    \label{eq:FC3}
F_\pm(\mathbf{x},\mathbf{x}_f,t)\hspace{-2mm}&=&\hspace{-2mm}0,\ \mathrm{for}\ t\uparrow+\infty, \\
&\mathrm{and} &\nonumber\\
    \label{eq:FC4}
    F_\pm(\mathbf{x},\mathbf{x}_f,t)\Bigg|_{z=z_f}\hspace{-3mm}&=&\hspace{-2mm}\delta(t)\delta(\mathbf{x}_H-\mathbf{x}_{H,f}),\\
    \label{eq:FC5}
    \hspace{-0.5cm}\left[\frac{\partial}{\partial z} \mp \sqrt{\frac{1}{c^2(z)}\frac{\partial^2}{\partial t^2} - \frac{\partial^2}{\partial x^2} - \frac{\partial^2}{\partial y^2}}\ \right]F_\pm(\mathbf{x},\mathbf{x}_f,t)\Bigg|_{z=z_f}\hspace{-3mm}&=&\hspace{-2mm}0,
\end{eqnarray}
where $\nabla^T=\left(\frac{\partial}{\partial x},\frac{\partial}{\partial y},\frac{\partial}{\partial z}\right)$, and the subscript $\pm$ of $F$ denotes respectively an down- or upgoing focusing solution at $z=z_f$. These conditions are illustrated in Figure~1. The initial and final conditions are required to eliminate a null-space containing waves that enter or leave the domain boundaries, without passing through the focal level (and therefore avoid constraint by the focusing conditions).

\begin{figure}[htp]
     \centering
     \begin{subfigure}[b]{0.5\textwidth}
         \centering
         \includegraphics[width=\textwidth]{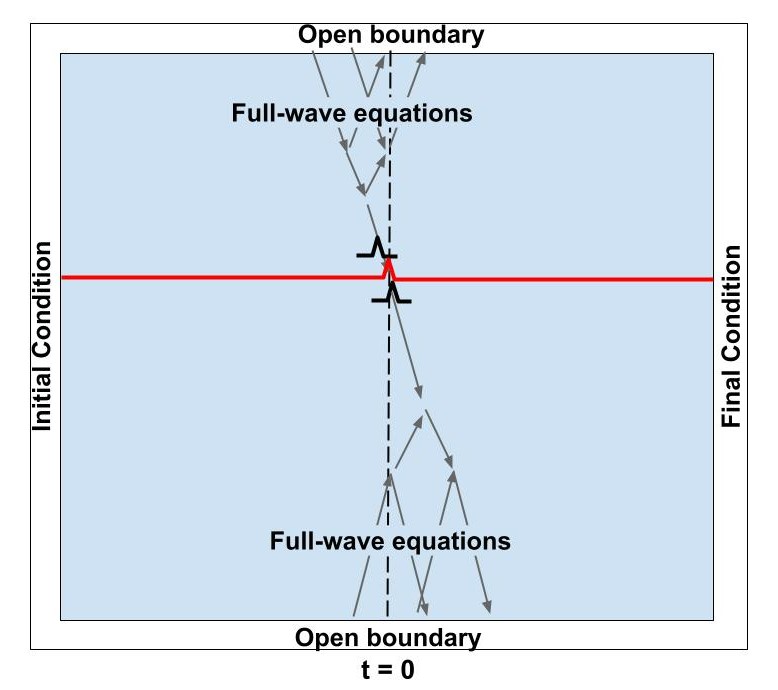}
         \label{fig:focusing-a}
     \end{subfigure}
     \begin{subfigure}[b]{0.4\textwidth}
         \centering
		\includegraphics[width = \textwidth]{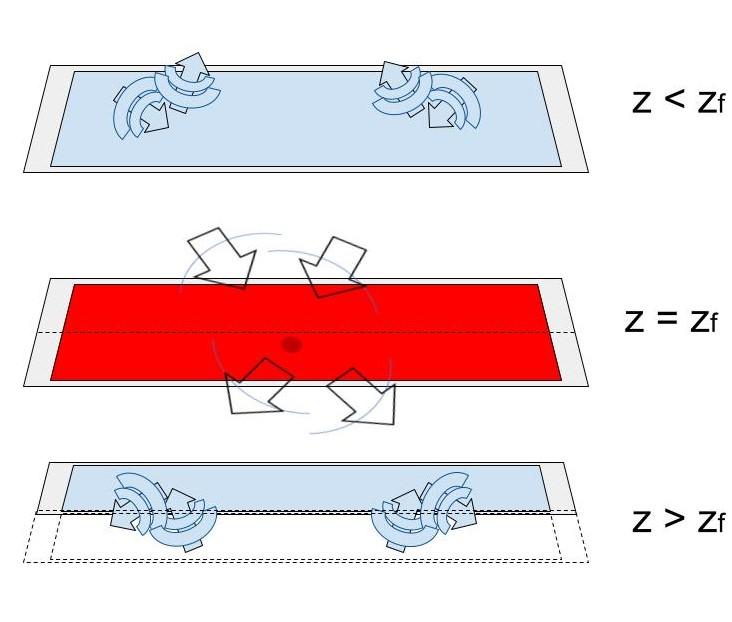}
		\label{fig:focusing-b}
     \end{subfigure}

        \caption{Conceptual illustration of a focusing wavefield (here a down-going focusing wavefield). (a) Focusing conditions at $z=z_f$ (the focal plane) consist of the condition that the wavefield forms an impulse at $\mathbf{x}=\mathbf{x}_{H,f}$ and obeys the one-way wave equation (for down-going waves). Ergo, there is no propagation on the focal plane. Above and below the focal plane there are super-positions of both up- and down-going waves. (b) Illustration of the domain's boundary, initial, final, and focusing conditions on the wavefield.}
        \label{fig:focusing}
\end{figure}

\subsection{Solving for focusing functions as a wavefield reconstruction problem.}
Similar as in wavefield reconstruction inversion for waveform inversion (\cite{leeuwen2013a}; \cite{vanLeeuwen16}), we combine the focusing conditions and the wave equation in a PDE constrained optimisation problem. The desired wavefield (here denoted with $\mathbf{p}$) and medium parameters are discretised on a finite Cartesian basis and all derivatives are approximated using finite difference stencils. A band-limited solution to the focusing function is found by imposing the focusing conditions using a band-limited focus signature in time, $s(t)$, and space, $s(r_H)$, where $r_H=\big|\mathbf{x}_{H}-\mathbf{x}_{H,f}\big|$, organised in $\mathbf{s}$. The wavefield solution can be solved as a wavefield reconstruction inversion problem with the following objectives:
\begin{eqnarray}
\label{eqn:fwri}
\mathbf{M}\mathbf{L}\mathbf{p} = \mathbf{0}\hspace{0.5cm}\mathrm{and}\hspace{0.5cm}\mathbf{K}\mathbf{p} = \mathbf{q},\hspace{0.5cm} \mathrm{where}\hspace{0.5cm}\mathbf{q}=\left[\begin{array}{c}\mathbf{s} \\
\hline
\mathbf{0}\end{array}\right].
\end{eqnarray}
Here $\mathbf{L}$ is the discretised wave equation operator, and $\mathbf{M}$ is a boundary exclusion operator. The matrix operator $\mathbf{K}$ is a structure with two blocks that (above) contains a sampling operator selecting the wavefield at the focal level (mapping $\mathbf{p}$ unto $\mathbf{s}$) and (below) contains the discretised one-way wave equation operator (mapping $\mathbf{p}$ unto $\mathbf{0}$). Strictly, $\mathbf{K}$ should also contain sampling operators for the initial and final conditions (mapping $\mathbf{p}$ unto $\mathbf{0}$), see below for further discussion.

These objectives can be solved by minimizing the following cost function:
\begin{eqnarray}
\label{eq:costTK}
C(\mathbf{p})=\big\Vert\mathbf{M}\mathbf{L}\mathbf{p}\big\Vert_2^2 + \big\Vert\mathbf{K}\mathbf{p}-\mathbf{q}\big\Vert_2^2 + \lambda\big\Vert\mathbf{J}\mathbf{p}\big\Vert_2^2,
\end{eqnarray}
where the third term applies zero-order Tikhonov regularisation to the solution. The wavefield solution at the focal level is omitted from the Tikhonov constraint using the masking operator $\mathbf{J}$. A Tikhonov constraint is redundant if the solution is well-posed (and lacks a null-space), but in other inverse problems it may be needed to resolve null-spaces or damp spurious energy in the solution if the solution is over-posed with conflicting fitting objectives. In this study we find the minimum of the cost function in Equation~\ref{eq:costTK} with the following equivalent least-squares solution:
\begin{eqnarray}
\label{eq:wfr_solution_TK}
\mathbf{p}=\left[\mathbf{L}^T\mathbf{M}^T\mathbf{M}\mathbf{L} + \mathbf{K}^T\mathbf{K} + \lambda \mathbf{J}^T\mathbf{J}\right]^{-1}\mathbf{K}^T\mathbf{q}.
\end{eqnarray}

\section{Example focusing fields in 1D}
We start by computing the solution of Equation~\ref{eq:wfr_solution_TK} in the $(z,t)$ domain and we solve for $\mathbf{p}$ with $\lambda=0$. In this 1D $(z,t)$ domain case, the matrix and vector elements in Equation~\ref{eqn:fwri} are constructed as:
\begin{eqnarray}
\mathbf{M}\mathbf{L}\mathbf{p}=\hspace{10cm}\\
\left[\begin{array}{cccccccccccc}
0 & 0 & 0 & \cdots & 0 & 0 & 0 & \cdots\\
a_{2} & 2v_{2}^2-w_{2} & b_{2} & \cdots & 0 & -v_{2}^2 & 0 & \ddots\\
0 & a_{3} & 2v_{3}^2-w_{2} & \cdots & 0 & 0 & -v_{3}^2 & \ddots\\
\vdots & \vdots & \vdots & \ddots & \vdots & \vdots & \vdots & \ddots \\
0 & 0 & 0 & \cdots & 0 & 0 & 0 & \ddots\\
0 & -v_{2}^2 & 0 & \cdots & a_{2} & 2v_{2}^2-w_{2}& b_{2} & \ddots\\
0 & 0 & -v_{3}^2 & \cdots & 0 & a_{3} & 2v_{3}^2-w_{2} & \ddots \\
\vdots & \ddots & \ddots & \ddots & \ddots & \ddots & \ddots & \ddots
\end{array}\right]
\left[\begin{array}{c}
p_{z_1,t_1}\\
p_{z_2,t_1}\\
p_{z_3,t_1}\\
\vdots\\
p_{z_1,t_2}\\
p_{z_2,t_2}\\
p_{z_3,t_2} \\
\vdots
\end{array}\right],\hspace{-0.75cm} \nonumber
\end{eqnarray}
\begin{eqnarray}
\mathbf{K}\mathbf{p}=\hspace{10.4cm}\\
\left[\begin{array}{cccccccccccccccc}
0 & 1 & 0 & \cdots & 0 & 0 & 0 & \cdots & 0 & 0 & 0 & \cdots\\
0 & 0 & 0 & \cdots & 0 & 1 & 0 & \cdots & 0 & 0 & 0 & \ddots \\
0 & 0 & 0 & \cdots & 0 & 0 & 0 & \cdots & 0 & 1 & 0 & \ddots \\
\vdots & \ddots & \ddots & \ddots & \ddots & \ddots & \ddots & \ddots & \ddots & \ddots & \ddots & \ddots \\
\hline
-1 & 0 & 1 & \cdots & 0 & \hspace{-1mm}\pm v_{z_f}\hspace{-3mm} & 0 & \cdots & 0 & 0 & 0 & \cdots \\
0 & \hspace{-1mm}\mp v_{z_f}\hspace{-3mm} & 0 & \cdots & -1 & 0 & 1 & \cdots & 0 & \hspace{-1mm}\pm v_{z_f}\hspace{-3mm} & 0 & \ddots\\
0 & 0 & 0 & \cdots & 0 & \hspace{-1mm}\mp v_{z_f}\hspace{-3mm} & 0 & \cdots & -1 & 0 & 1 & \ddots\\
\vdots & \ddots & \ddots & \ddots & \ddots & \ddots & \ddots & \ddots & \ddots & \ddots & \ddots & \ddots
\end{array}\right]
\left[\begin{array}{c}
p_{z_f-1,t_1}\\
p_{z_f,t_1}\\
p_{z_f+1,t_1}\\
\vdots\\
p_{z_f-1,t_2}\\
p_{z_f,t_2}\\
p_{z_f+1,t_2}\\
\vdots\\
p_{z_f-1,t_3}\\
p_{z_f,t_3}\\
p_{z_f+1,t_3}\\
\vdots
\end{array}\right]\nonumber\hspace{-0.8cm}
\end{eqnarray}
and
\begin{eqnarray}
\mathbf{q}=\Big[\begin{array}{cccc|cccc}s(t_1) & s(t_2) & s(t_3) & \cdots & 0 & 0 & 0 & \cdots\end{array}\Big]^T,
\end{eqnarray}
where
\begin{eqnarray}\label{eq:abw}
a_{i} = \frac{1}{2}\left(1+\frac{\rho(z_i)}{\rho(z_{i-1})}\right),\hspace{0.5cm}
b_{i} = \frac{1}{2}\left(1+\frac{\rho(z_i)}{\rho(z_{i+1})}\right),\hspace{0.5cm}
w_{i} = a_{i} + b_{i},
\end{eqnarray}
and $v_{z_i}=\frac{\Delta z}{\Delta t}\frac{1}{c(z_i)}$. The initial and final conditions are not explicitly imposed. Instead, we implement wrap-around in time. This ensures that the domain is effectively closed and all energy that enters the medium will pass the focal level and should therefore be in agreement with the focusing constraints.

\begin{figure}[H]
	\begin{center}
		\includegraphics[width = \textwidth]{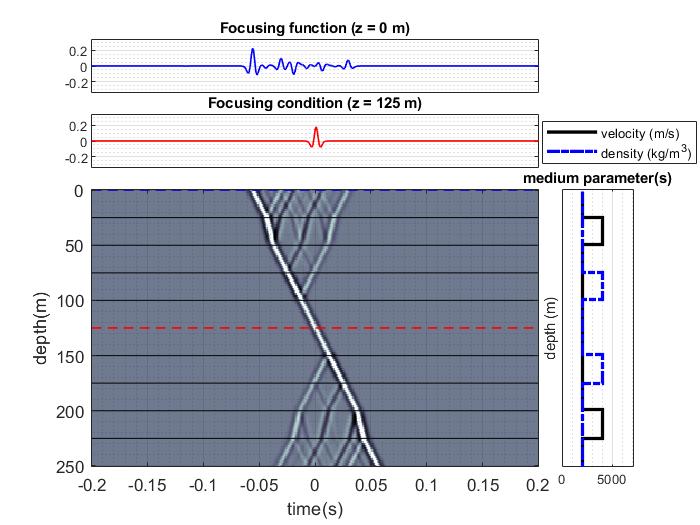}
		\caption{Down-going focusing function 1D solution solved in $(z,t)$. Right hand panel contains the medium structure. Top panel contains two time traces at respectively the top of the solution domain and the focal level.}
		\label{fig:wfr_tz_1d}
	\end{center}
\end{figure}

An example is shown in Figure~\ref{fig:wfr_tz_1d}. The spatial domain of the solution is $250$~m deep, the focal level at a depth of $125$~m, and the medium contains both velocity and density contrasts (see the right panel in Figure~\ref{fig:wfr_tz_1d}).

\section{Example Focusing fields in 1.5D}
We continue by computing the solution of Equation~\ref{eq:wfr_solution_TK} in the $(k_x,z,\omega)$ domain. Note that in this case, imposing explicit initial and final conditions would require all frequencies to be solved simultaneously. The null-space associated with the absence of the initial and final condition vanishes because all energy that enters the domain must pass through the focal level. In the 1.5D $(k_x,z,\omega)$ domain case, the matrix and vector elements in Equation~\ref{eqn:fwri} are constructed as:
\begin{eqnarray}
\hspace{-2mm}\mathbf{M}\mathbf{L}\mathbf{p}=\hspace{9.8cm}\\
\hspace{-2mm}\left[\begin{array}{ccccccccccc}
0 & 0 & 0 & \cdots & 0 & 0 & 0 \\
a_{2} & \hspace{-1mm}h_{2}^2-w_{2} \hspace{-2mm} & b_{2} & \cdots & 0 & 0 & 0 \\
0 & a_{3} & \hspace{-2mm} h_{3}^2-w_{3} \hspace{-2mm} & \cdots & 0 & 0 & 0 \\
\vdots & \vdots & \vdots & \ddots & \vdots & \vdots & \vdots \\
0 & 0 & 0 & \cdots & \hspace{-1mm}h_{N-2}^2-w_{N-2} \hspace{-3mm} & b_{N-2} & 0 \\
0 & 0 & 0 & \cdots & a_{N-1} & \hspace{-2mm} h_{N-1}^2-w_{N-1}  & b_{N-1} \\
0 & 0 & 0 & \cdots & 0 & 0 & 0 \\
\end{array}\right]
\left[\begin{array}{c}
p_{z_1,\omega,k_x}\\
p_{z_2,\omega,k_x}\\
p_{z_3,\omega,k_x}\\
\vdots\\
p_{z_{N-2},\omega,k_x}\\
p_{z_{N-1},\omega,k_x}\\
p_{z_N,\omega,k_x}
\end{array}\right],\hspace{-1cm}\nonumber
\end{eqnarray}
\begin{eqnarray}
\mathbf{K}\mathbf{p}=\left[\begin{array}{ccccccccccc}
\renewcommand*{\arraystretch}{2.0}
0 & \cdots& 0  & 1 & 0 & \cdots & 0 \\
\hline
0 & \cdots& 1 & \pm 2\,i\,h_{z_f} & -1 & \cdots & 0
\end{array}\right]
\left[\begin{array}{c}
p_{z_1,\omega,k_x} \\
\cdots \\
p_{z-1,\omega,k_x} \\
p_{z,\omega,k_x} \\
p_{z+1,\omega,k_x} \\
\cdots \\
p_{z_N,\omega,k_x}
\end{array}\right],
\end{eqnarray}
and
\begin{eqnarray}
\mathbf{q}=\left[\begin{array}{c}
s(\omega)s(k_x)\\
\hline
0
\end{array}\right],
\end{eqnarray}
where $a_i, b_i$, and $w_i$ are given in Equation~\ref{eq:abw} and $h_i=\Delta z \sqrt{\frac{\omega^2}{c^2(z_i)}-k_x^2}$. Note that, conversely to the 1D case, here the minimum of the cost function in Equation~\ref{eq:costTK} is performed independently for each $f,k_x$ pair, and the combined set of resulting $\mathbf{p}$ vectors are transformed back to time and horizontal space using the 2D Fourier transform.

\begin{figure}[htp]
     \centering
     \begin{subfigure}[b]{0.49\textwidth}
         \centering
         \includegraphics[width=\textwidth]{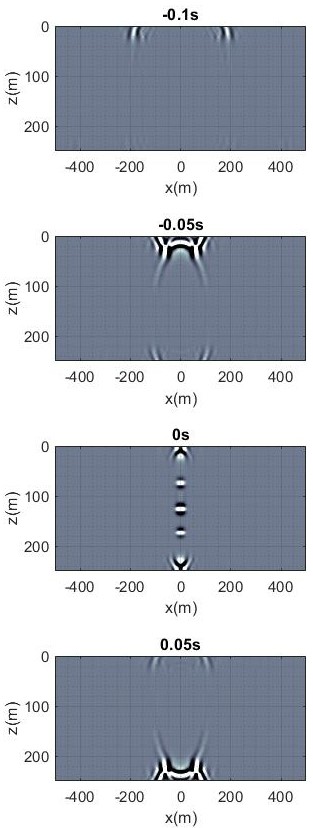}
         \caption{}
         \label{fig:FF1p5D-a}
     \end{subfigure}
     \begin{subfigure}[b]{0.49\textwidth}
         \centering
		\includegraphics[width=\textwidth]{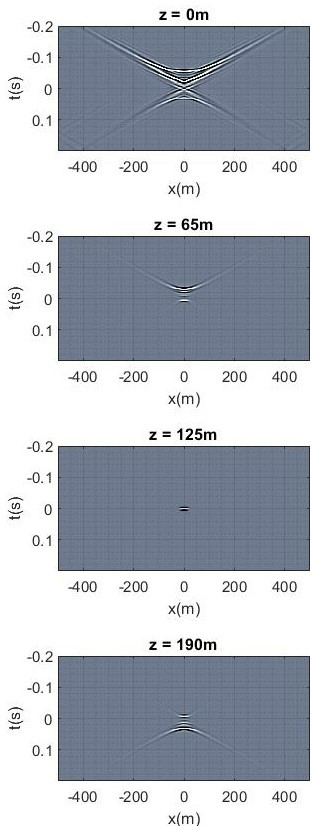}
		\caption{}
		\label{fig:FF1p5D-b}
     \end{subfigure}
        \caption{1.5D down-going focusing function solution solved in $(k_x,\omega,z)$. Solved using $\lambda = 10^{-8}$. (a) Cross-sections in the  $(x,z)$ plane. (b) Cross-sections in the  $(x,t)$ plane.}
        \label{fig:FF1p5D}
\end{figure}

\begin{figure}[htp]
     \centering
     \begin{subfigure}[b]{0.49\textwidth}
         \centering
         \includegraphics[width=\textwidth]{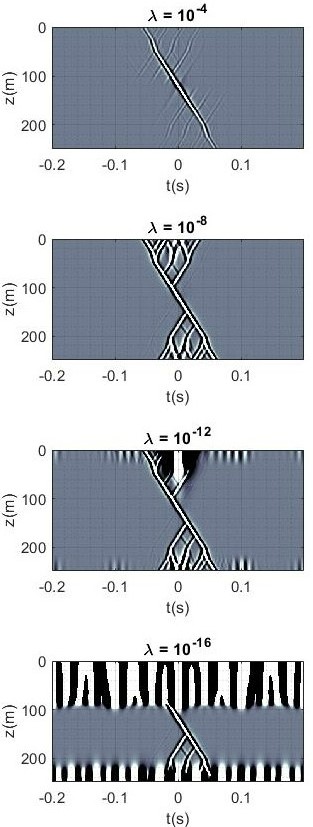}
         \caption{}
         \label{fig:FF1p5Dlambda-a}
     \end{subfigure}
     \begin{subfigure}[b]{0.49\textwidth}
         \centering
		\includegraphics[width = \textwidth]{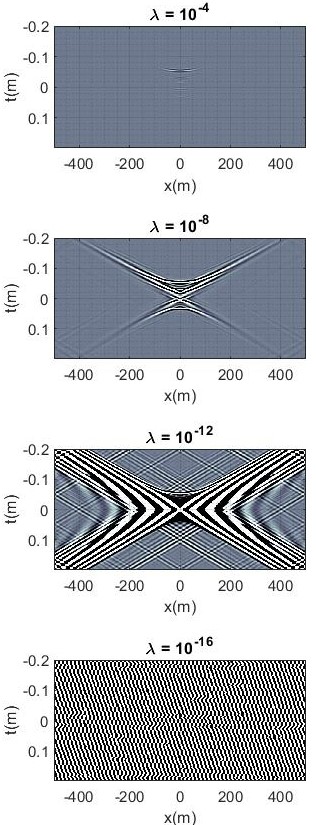}
		\caption{}
		\label{fig:FF1p5Dlambda-b}
     \end{subfigure}
        \caption{
        Down-going focusing function 1.5D solution solved in $(k_x,z,\omega)$ for different regularisation strengths. From left to right $\lambda=10^{-4},\ 10^{-8},\ 10^{-12},\ 10^{-16}$ (a) Cross-sections in the $(t,z)$ plane. (b) Cross-sections in the $(x,t)$ plane.}
        \label{fig:FF1p5Dlambda}
\end{figure}

An example is shown in Figure~\ref{fig:FF1p5D}. The spatial domain of the solution is $250$~m deep, $1000$~m wide, the focal level at a depth of $125$~m, and the medium contains both velocity and density contrasts (see the right panel in Figure~\ref{fig:wfr_tz_1d}). The focusing function solution in Figure~\ref{fig:FF1p5D} was computed using Tikhonov regularisation ($\lambda = 10^{-8}$). We explore the effect of the regularisation strength on the focusing function solutions in Figure~\ref{fig:FF1p5Dlambda}.

\section{Discussion}
\begin{figure}[htp]
     \centering
     \begin{subfigure}[b]{1.0\textwidth}
         \centering
 		\includegraphics[width = 0.8\textwidth]{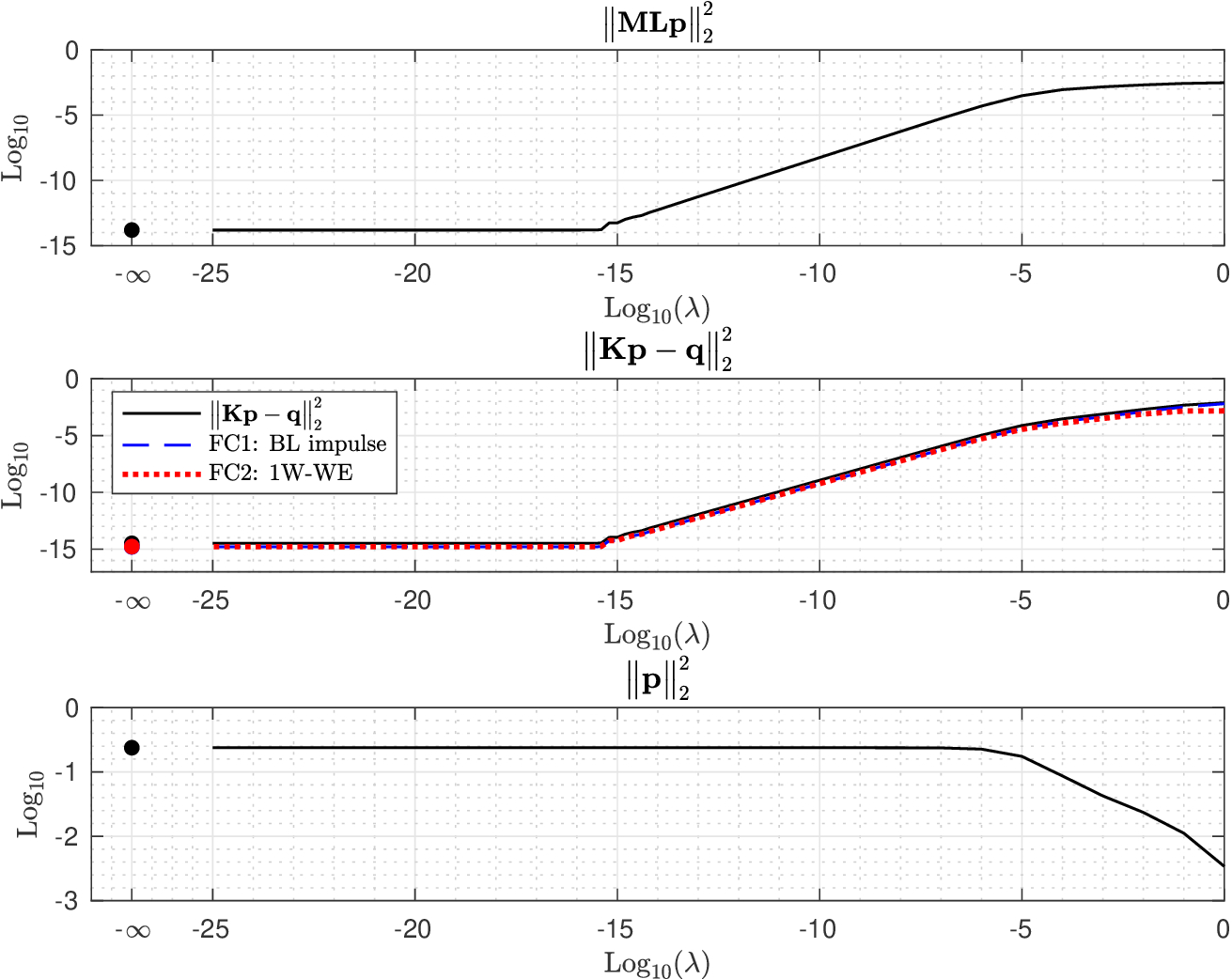}
         \caption{}
         \label{fig:cost_k0}
     \end{subfigure}
     \begin{subfigure}[b]{1.0\textwidth}
        \centering
		\includegraphics[width = 0.8\textwidth]{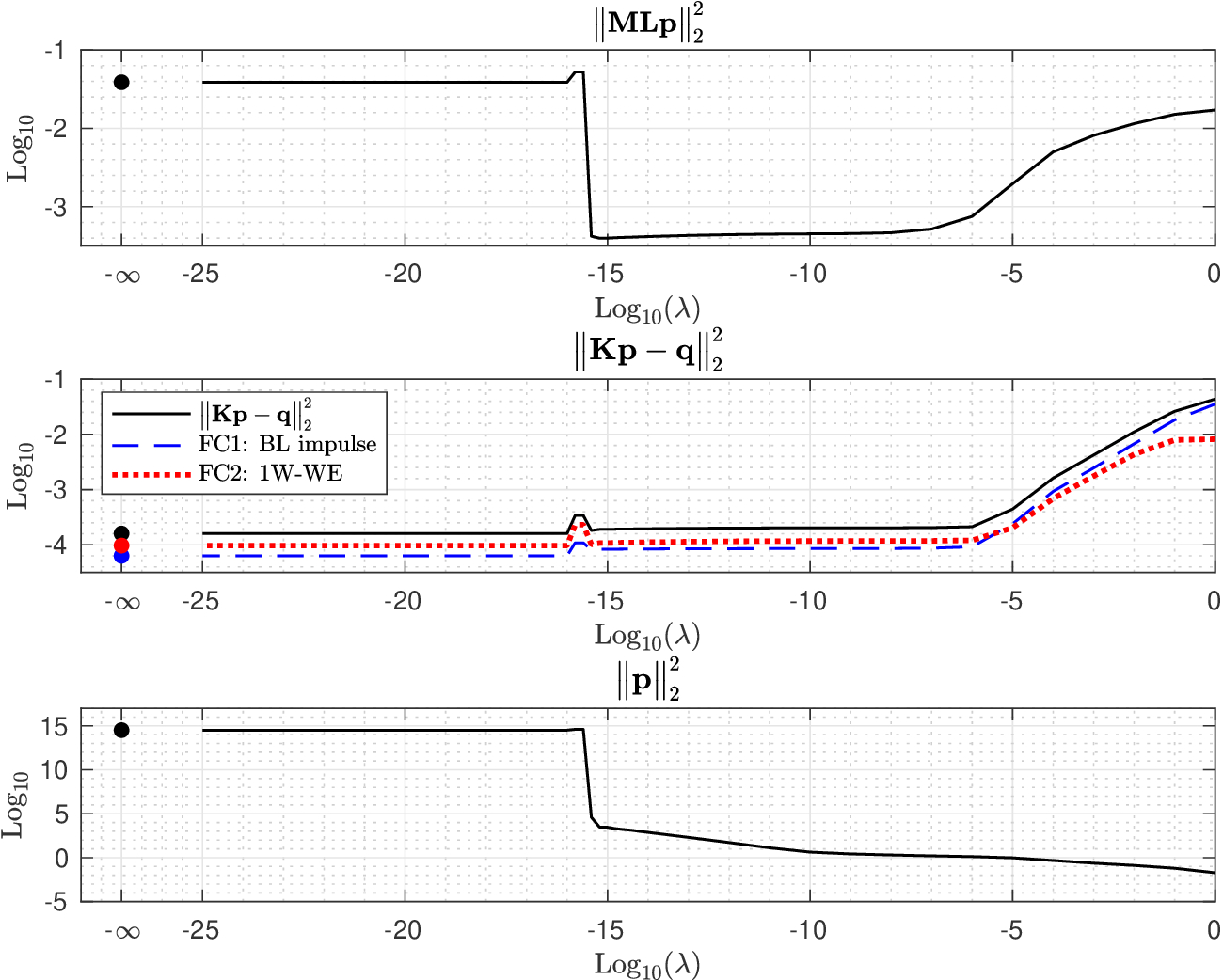}
		\caption{}
		\label{fig:cost}
     \end{subfigure}

     \caption{(a) Cost function terms (in Equation~\ref{eq:costTK}) at the inverted focusing function solution in 1.5D as a function of Tikhonov regularisation strength, when solved for $k_x=0$ only. (b) Cost function terms (in Equation~\ref{eq:costTK})at the inverted focusing function solution in 1.5D as a function of Tikhonov regularisation strength, solved and summed for all $k_x$.}
        \label{fig:costs}
\end{figure}

The focusing solutions in higher dimensions ($>1D$) contain very strong evanescent energy. We observe that this evanescent energy can be damped using Tikhonov regularisation. The question arises to what is the optimum regularisation strength. Figure~\ref{fig:costs} contains an analysis of the magnitude of the various terms in the cost function of Equation~\ref{eq:costTK}, as a function of the Tikhonov regularisation strength ($\lambda$). FC1 and FC2 denote the individual contributions of respectively focusing conditions~\ref{eq:FC4} and~\ref{eq:FC5} in the cost function term containing the focusing conditions. Notice that there is no regularisation strength for which both the focusing conditions and the wave equation constraint are at their minimum (neither for $\lambda=0$). This appears to be a situation for solutions in higher dimensions ($>1D$). We tested the same solutions when incorporating the $k_x=0$ component only and find that both the focusing conditions and the wave equation constraint are at their minimum with vanishing regularisation strength (including when $\lambda=0$). These examples indicate that exact solutions to the focusing function definition (Equations~\ref{eq:FC1}~to~\ref{eq:FC5}) in higher dimensions ($>1D$) are challenging to compute via numerical means.

\section{Conclusion}
The focusing function can be computed as the solution to a wavefield reconstruction inversion problem. We find and show that solutions in higher dimensions are dominated by strong evanescent energy. This energy has been hypothesised to exist and is considered a critical part of a focusing function in higher dimensions that ensures perfect focusing at depth. Their presence in the inverse solution can be damped using Tikhonov regularisation. Posing the Marchenko focusing function the solution to a PDE constrained optimisation problem may allow for novel seismic waveform inversion schemes.

\section{Acknowledgements}
RFH is grateful for financial support from the Indonesian Endowment Fund for Education/Lembaga Pengelola Dana Pendidikan (LPDP) and thanks the Leeds Turing scheme, KAUST, and M. Ravasi for supporting his visit to KAUST.

\printbibliography

\end{document}